\documentclass[aps,prd,groupedaddress,nofootinbib]{revtex4-2}
\usepackage{amsmath,amssymb,amsfonts}
\usepackage{graphicx}
\usepackage{hyperref}
\usepackage{physics}
\usepackage{bbm}

\begin{document}

\title{Resonance-Protected Pointer States: Stationary-Phase Analysis of Entanglement in an Environment-Coupled Two-Spin System}

\author{Kentaro Urasaki}
\email{urasakikentaro@gmail.com}
\affiliation{Tokyo, Japan}

\date{\today}

\begin{abstract}
In this manuscript, we analyze the dynamics of a model in which two spin-1/2 systems are each
coupled noncommutingly to a static environmental field and to a self-Hamiltonian, while also
being mutually coupled through a $\sigma_y^{(1)}\sigma_y^{(2)}$ interaction. By constructing an
exact solution that exploits parity symmetry, and by applying the stationary-phase
approximation (the saddle-point method) in the continuum limit of the environmental field, we
show that coherence associated with the ordinary isolated stationary point decays as $t^{-1}$,
whereas, under the resonance condition at which the local effective fields of the two spins
cancel each other, only the coherence between a pair of maximally entangled, Bell-type states—
the eigenbasis of the inter-system coupling—survives, decaying via the anomalously slow power
law $t^{-1/2}$. Through a comparison with the pointer-basis theory of W. H. Zurek and coworkers
(the 1981 and 2005 models), we further reveal a mechanism not reducible to local coupling: in
this system, the pointer observable is determined not by the individual system–environment
couplings but by a statistical resonance condition between the two environments.
\end{abstract}


\maketitle


\section{Introduction}\label{sec:intro}
With the advancement of quantum technologies—including quantum communication, quantum sensing,
and quantum computation—the distance between the quantum-mechanical world and the macroscopic
world of everyday experience has been steadily narrowing. In this context, quantum decoherence,
which addresses the boundary between quantum-mechanical description and macroscopic experience,
has been vigorously analyzed from a variety of models and perspectives, and our understanding of
it has steadily deepened~\cite{zeh1970, zurek1981, caldeira1983, joos1985}.

However, since real environments consist of countless degrees of freedom, and the interaction
structure woven between the system and its environment is extremely complex, it is difficult to
predict the emergence of a stable classical world from simple models alone. This, in turn,
harbors a profound question concerning how one transitions from the known microscopic physical
laws to our epistemological level of description. In response to this problem, Zurek and
coworkers, for example, introduced information-theoretic criteria such as Quantum Darwinism,
which regards the environment as a channel of information transmission, and have made important
progress in recent years toward explaining the stability and objectivity of the classical states
we experience~\cite{zurek2009, korbicz2021}.

In particular, identifying the ``pointer states'' that can exist stably under interaction with
the environment during the decoherence process is, in itself, an essential problem, in the sense
that it determines which physical quantities are selected as classical states. Pointer states
also constitute a concept bridging fundamental physical laws with information theory and
epistemology; from this perspective, efforts to elucidate the emergence mechanism of pointer
states themselves through bottom-up approaches are equally important.

Through the accumulation of these aforementioned pioneering studies, the general principle that
local system–environment coupling structures determine pointer states is becoming established in
the current literature~\cite{zurek1981, caldeira1983, joos1985}. However, a complete consensus
regarding the detailed mechanism has yet to be reached~\cite{schlosshauer2004}.

In this paper, we further extend the most representative bit-by-bit model to a higher hierarchical
level and, by analyzing a regime in which an exact solution can be obtained, report a new
mechanism whereby ``resonance''—a statistical condition between the two environments—determines
the two-body pointer ``observable.'' This mechanism yields a two-body entangled state that is
protected, and thus rendered longer-lived than usual, against local environmental noise.

\subsection{The Concept of Pointer States and Prototypical Models}
The study of quantum decoherence has been organized around several systematic classes of models.
A representative example is the bit-by-bit model introduced by Zurek et al.~\cite{zurek1981, 2005}.
In this model, the entanglement between the central system and each bit constituting the environment
can be tracked in considerable detail, in a relatively exact manner.

Consider the total system consisting of a central spin coupled to $N$ environmental bits
$E_i$ ($i=1,\dots,N$). The total Hamiltonian takes the form
\begin{equation}
  h = \Delta\sigma_x+\frac{1}{2}\sigma_z\otimes\sum_{i=1}^Ng_i\sigma_z^{i}\equiv\Delta\sigma_x+\frac{1}{2}\sigma_z\otimes E,
\end{equation}
where $\sigma_z^{i}$ denotes the Pauli operator of the $i$-th environmental bit $E_i$,
$g_i$ is the corresponding coupling constant, and
$E \equiv \sum_i g_i \sigma_z^i$ is the collective operator that aggregates the environmental
degrees of freedom.

Since the environment possesses no self-Hamiltonian, the product basis of the $N$ bits,
$\ket{n} = \bigotimes_{i=1}^N \ket{\phi_i^{(n_i)}}$ ($n = 0,\dots,2^N-1$, with $n_i$ the $i$-th
binary digit of $n$), built from the $z$-eigenstates of each environmental spin, commutes with
the total Hamiltonian. Consequently, the time-evolution operator can be written exactly in
block-diagonal form with respect to each environmental configuration $n$:
\begin{eqnarray}
{\mathcal U}(t)=e^{-iht/\hbar}\\
=\prod^{2^N-1}_{n=0}U_{B(n)}(t)\otimes\ket{n}\bra{n},
\end{eqnarray}
which reduces the problem to solving
\begin{equation}
h(n) = \Delta \sigma_x+ B(n)\,\sigma_z
\end{equation}
for each subspace, with a different static external field applied in each case.
Here $B(n)$ is the static effective field exerted on the central spin by the $n$-th eigenstate,
and $U_{B(n)}(t)$ is the effective time evolution of the central system when the environment is
in configuration $n$.

The prototypical bit-by-bit model introduced by Zurek~\cite{zurek1981} corresponds to the special
case $\Delta = 0$ in this general form, i.e., the limit in which the central system has no
self-Hamiltonian. In this case, $h$ commutes with $\sigma_z$, so the pointer basis of the central
system is trivially given by the eigenstates $\ket{\mu}$ ($\mu = 0,1$) of $\sigma_z$, and the
resulting decoherence is pure dephasing—that is, phase relaxation unaccompanied by energy
dissipation.

Indeed, if the initial state of the central system is taken as
$\ket{\psi_S(0)} = c_0\ket{0} + c_1\ket{1}$, the off-diagonal element of the reduced density
matrix is given by
\begin{equation}
  \rho_{01}(t) = c_0 c_1^{*}\, r(t),
  \qquad
  r(t) = \prod_{i=1}^{N} \cos(g_i t).
  \label{eq:decoherence-factor}
\end{equation}
In this case, the decoherence factor $r(t)$ becomes a Gaussian function of time in the continuum
limit, decaying rapidly to $0$ as $t\to\infty$.

On the other hand, the case $\Delta \ne 0$ yields a more realistic model~\cite{zurek2005}.
Since there is again no interaction among the environmental bits, the same block-diagonalization
as above remains possible, and the model can be solved exactly (though some assumption on the
environmental state is required). In this case, the decoherence factor generally exhibits
exponential decay. Importantly, in the intermediate parameter regime—where the self-Hamiltonian
and the interaction are of comparable magnitude—the eigenbasis of the interaction Hamiltonian 
cannot be identified outright as the pointer state. A competition arises between the energy
eigenstates and the states that diagonalize the interaction Hamiltonian, and which of the two is
actually selected as the pointer state depends on their relative magnitudes.

\section{Theoretical Setup: Hierarchical Extension to Macroscopic Systems}\label{sec:setup}

\subsection{Construction of the Two-Spin Model}
To connect this discussion to macroscopic systems, it is natural to introduce a further,
higher level of hierarchy. We extend the central spin discussed above to two spins and
consider a model in which these two spins interact with each other.
Each spin interacts with its own environment,
\begin{align}
  h^{(1)}= \Delta_1 \sigma_x^{(1)} 
  +\frac{1}{2}\sigma_z^{(1)}\otimes E^{(1)} \\
  h^{(2)}= \Delta_2 \sigma_{x'}^{(2)} 
  +\frac{1}{2}\sigma_{z'}^{(2)}\otimes E^{(2)}. 
\label{eq:hi}
\end{align}
The total Hamiltonian of the full system is
\begin{equation}
  H= h^{(1)}+h^{(2)}+ g\,\sigma_y^{(1)}\sigma_y^{(2)}.
\label{eq:H}
\end{equation}
Here, the self-Hamiltonian of each spin does not commute with its interaction with the
environment, and the two spins are taken to have the same type of coupling to their
respective environments. Note that $x, z$ and $x', z'$ can be chosen independently.

After mapping onto the environmental configurations, the Hamiltonian is given by
\begin{equation}
H(n,m) = \Delta_1 \sigma_x^{(1)} + \Delta_2 \sigma_{x'}^{(2)}
+ B_1(n)\,\sigma_z^{(1)} + B_2(m)\,\sigma_{z'}^{(2)} + g\,\sigma_y^{(1)}\sigma_y^{(2)}.
\end{equation}

Remarkably, this model can also be solved exactly (although some assumption on the nature
of the environmental states is required in order to compute explicit results). Here,
$B_1(n)$ and $B_2(m)$ are the static effective fields depending on the discrete
environmental-state labels $n$ and $m$, respectively.

\subsection{Basis Rotation and the Exact Solution}
For each spin, we combine the local field components and perform a basis rotation such
that the quantization axis is aligned with the direction of the resulting field,
\begin{align}
\Omega_1(n) &= \sqrt{\Delta_1^2 + B_1(n)^2}, \quad \tan\theta_1(n) = \frac{\Delta_1}{B_1(n)}, \\
\Omega_2(m) &= \sqrt{\Delta_2^2 + B_2(m)^2}, \quad \tan\theta_2(m) = \frac{\Delta_2}{B_2(m)}.
\end{align}
Denoting the rotated spin axes by $\xi(1)$ and $\zeta(2)$, respectively, the Hamiltonian
reduces to the simple form
\begin{equation}
H(n,m) = \Omega_1(n) \sigma_\xi^{(1)} + \Omega_2(m) \sigma_\zeta^{(2)} + g \sigma_y^{(1)} \sigma_y^{(2)}.
\end{equation}

The Hamiltonian commutes with the parity operator $\Pi = \sigma_\xi^{(1)} \sigma_\zeta^{(2)}$
(i.e., $[H, \Pi] = 0$), so the full four-dimensional Hilbert space decomposes completely
into two $2 \times 2$ blocks labeled by $\Pi = \pm 1$. The eigenenergies in each block are
obtained exactly as
\begin{align}
E^+_{\pm}(n, m) &= \pm \sqrt{(\Omega_1(n) + \Omega_2(m))^2 + g^2} \quad (\Pi = +1), \\
E^-_{\pm}(n, m) &= \pm \sqrt{(\Omega_1(n) - \Omega_2(m))^2 + g^2} \quad (\Pi = -1).
\end{align}

\section{Static Gaussian Environmental Fields and the Stationary-Phase Approximation}\label{sec:stationary-phase}

\subsection{Reduced Density Matrix: Continuum Approximation for the Environment and the Independent Adiabatic Approximation}

We assume static environmental bits. Under this assumption, given the initial
environmental distributions $P(n)$ and $P(m)$, the reduced density matrix can be written as
\begin{equation}
\rho_S(t) = \sum_{n,m} P(n)\,P(m)\; U(n,m,t)\,\rho_S(0)\,U(n,m,t)^\dagger.
\end{equation}
When the number of environmental degrees of freedom is large enough that $B_1(n)$ and
$B_2(m)$ can be treated as continuous variables, we replace the sums by integrals under
the Gaussian approximations $P(n)\to \mathcal{N}(0,\sigma_1^2)$ and
$P(m)\to\mathcal{N}(0,\sigma_2^2)$, obtaining the integral representation
\begin{equation}
\rho_S(t) = \int dB_1\, dB_2\;
\frac{e^{-B_1^2/2\sigma_1^2}}{\sqrt{2\pi}\,\sigma_1}\,
\frac{e^{-B_2^2/2\sigma_2^2}}{\sqrt{2\pi}\,\sigma_2}\;
U(B_1,B_2,t)\,\rho_S(0)\,U(B_1,B_2,t)^\dagger,
\end{equation}
where $\Omega_1=\Omega_1(B_1)=\sqrt{\Delta_1^2+B_1^2}$ and
$\Omega_2=\Omega_2(B_2)=\sqrt{\Delta_2^2+B_2^2}$.

\noindent
\textbf{Remark:} Care is required in the case of a shared bath in which $z$ and $z'$ do
not coincide: in this situation, the environmental configuration labels $n$ (the degree of
freedom to which spin 1 couples) and $m$ (the degree of freedom to which spin 2 couples)
originate from noncommuting readouts of the same bath, and thus cannot strictly be
regarded as independent variables. In such a case, the analysis presented in this paper
corresponds to an approximation in which this noncommutativity is neglected within the
accuracy of the adiabatic approximation, treating $n$ and $m$ as independent (slowly
varying) classical variables. A quantitative assessment of this correlation (or a precise
characterization of the adiabatic conditions under which it can be neglected) is left for
future work.

\subsection{Application of the Stationary-Phase Approximation and the Trivial Stationary Point}

\subsubsection{Spectral Decomposition and Superposition of Phases}
Performing a spectral decomposition of $U_p(n,m,t)$ ($p=+,-$) in each block using the
projectors $\Pi_p^{(s)}=P_p(\mathbbm{1}+sH_p/E_p)/2$ ($s=\pm1$) associated with the
eigenvalues $\pm E_p$, we obtain
\begin{equation}
U(B_1,B_2,t) = \sum_{p=\pm}\sum_{s=\pm1} e^{-isE_p(B_1,B_2)t}\,\Pi_p^{(s)}(B_1,B_2).
\end{equation}
Substituting this into the integral representation, $\rho_S(t)$ decomposes into a
superposition of phases,
\begin{equation}
\rho_S(t) = \sum_{p,p'}\sum_{s,s'} \int dB_1 dB_2\, P(B_1)P(B_2)\;
e^{-i\phi_{p,s;p',s'}(B_1,B_2)\,t}\;
\Pi_p^{(s)}\,\rho_S(0)\,\Pi_{p'}^{(s')},
\end{equation}
where $\phi_{p,s;p',s'}=sE_p-s'E_{p'}$.

\subsubsection{Non-oscillatory terms ($p=p',\ s=s'$)} For these terms, $\phi=0$ and the
contribution is constant, independent of $t$. This yields the diagonal ensemble (the
pointer decomposition) that survives as $t\to\infty$,
\begin{equation}
\rho_S(\infty) = \sum_{p,s}\int dB_1dB_2\,P(B_1)P(B_2)\,\Pi_p^{(s)}\rho_S(0)\Pi_p^{(s)}.
\end{equation}
The remaining terms, for which $\phi\neq0$, are the ``coherence'' terms, which we evaluate
below using the stationary-phase method.

\subsubsection{The isolated stationary point: $B_1=B_2=0$}
For the terms with equal parity and opposite sign ($p=p'$, $s=-s'$, with phase
$\phi=2sE_p$), one has
\begin{equation}
\frac{\partial E_p}{\partial B_1}\propto \frac{B_1}{\Omega_1},\qquad
\frac{\partial E_p}{\partial B_2}\propto \frac{B_2}{\Omega_2},
\end{equation}
so that $B_1=B_2=0$ is always a stationary point, for both $p=\pm$. Expanding $E_+$
around this point gives
\begin{equation}
E_+(B_1,B_2)\approx E_+^{(0)} + \frac{\Delta_1+\Delta_2}{2E_+^{(0)}\Delta_1}B_1^2 + \frac{\Delta_1+\Delta_2}{2E_+^{(0)}\Delta_2}B_2^2,
\qquad E_+^{(0)}\equiv E_+(0,0)=\sqrt{(\Delta_1+\Delta_2)^2+g^2},
\end{equation}
which is a positive-definite quadratic form. Applying the standard two-dimensional
stationary-phase formula, the contribution of this term decays as
\begin{equation}
\sim \frac{1}{t}\, e^{\,2iE_+^{(0)}t}
\end{equation}
(with an amplitude given by a finite coefficient determined by $P(0)^2$ and the Hessian).
Similarly, on the $E_-$ side, provided $\Delta_1\neq\Delta_2$ (i.e., in the generic case
away from the resonance manifold), this point is an isolated stationary point, giving the
same $\sim 1/t$ decay at frequency $2E_-^{(0)}=2\sqrt{(\Delta_1-\Delta_2)^2+g^2}$.

\subsection{Emergence of a Nontrivial Stationary Region (Hyperbolic-Type Saddle)}
In addition to this isolated point, $E_-$ possesses another, qualitatively distinct set of
stationary points. Examining the condition $\nabla \phi = 0$ for the phase function
$\phi(B_1, B_2) = - E(B_1, B_2) t / \hbar$ to have vanishing gradient, one finds that, for the
energy solution in the $\Pi = -1$ block, the stationarity condition is satisfied at every point
on the one-dimensional manifold (a family of hyperbolas) in $(B_1, B_2)$ space defined by the
condition that the composite Rabi frequencies coincide,
\begin{equation}
\Omega_1(B_1) = \Omega_2(B_2) \implies \sqrt{\Delta_1^2 + B_1^2} = \sqrt{\Delta_2^2 + B_2^2},
\end{equation}
that is,
\begin{equation}
B_2^2 - B_1^2 = \Delta_1^2 - \Delta_2^2.
\end{equation}

Since
\begin{equation}
\frac{\partial E_-}{\partial B_1}\propto(\Omega_1-\Omega_2)\frac{B_1}{\Omega_1},\qquad
\frac{\partial E_-}{\partial B_2}\propto-(\Omega_1-\Omega_2)\frac{B_2}{\Omega_2},
\end{equation}
\emph{both partial derivatives vanish identically} on the curve $\Omega_1(B_1)=\Omega_2(B_2)$—so
this is not an isolated point but a one-dimensional stationary curve (a degenerate set of
stationary points).

Introducing the coordinate along the curve, $\Omega_r\equiv\Omega_1=\Omega_2$, and the transverse
coordinate $u\equiv\Omega_1-\Omega_2$, we have $E_-(u)=\sqrt{u^2+g^2}\approx g+u^2/(2g)$. Using
the Jacobian $dB_1dB_2=\dfrac{\Omega_1\Omega_2}{B_1B_2}\,du\,d\Omega_r$, and defining
\begin{equation}
D(\Omega_r) \;\equiv\; \frac{\Omega_1 \Omega_2}{B_1 B_2}\bigg|_{\Omega_1=\Omega_2=\Omega_r}
= \frac{\Omega_r^2}{\sqrt{(\Omega_r^2-\Delta_1^2)(\Omega_r^2-\Delta_2^2)}},
\qquad \Omega_r > \max(\Delta_1,\Delta_2)
\end{equation}
(the condition $\Omega_r>\max(\Delta_1,\Delta_2)$ being required for a real resonance curve to
exist), the term with $p=p'=-,\ s=-s'$ (phase $\phi=2E_-(u)t$) becomes
\begin{equation}
C_-(t) \equiv \int\!\!\int dB_1dB_2\, P(B_1)P(B_2)\, e^{-i\,2E_-(u)\,t}
\;\approx\; \int d\Omega_r\, D(\Omega_r)\, P\big(B_1(\Omega_r)\big) P\big(B_2(\Omega_r)\big)
\int du\, e^{-i\left(2g+\frac{u^2}{g}\right)t}.
\end{equation}

The transverse Fresnel integral can be evaluated exactly,
\begin{equation}
\int du\, e^{-i u^2 t/g} = \sqrt{\frac{\pi g}{t}}\; e^{-i\pi/4},
\end{equation}
giving
\begin{equation}
C_-(t) \;\approx\; \mathcal{A}\; t^{-1/2}\, e^{-i(2gt+\pi/4)},
\qquad
\mathcal{A} \equiv \sqrt{\pi g}\int d\Omega_r\, D(\Omega_r)\, P\big(B_1(\Omega_r)\big) P\big(B_2(\Omega_r)\big).
\end{equation}
As a result, we obtain
\begin{equation}
\boxed{\ \text{Contribution of the resonance manifold}\ \sim\ \frac{1}{\sqrt{t}}\,e^{\,2igt}\ },
\end{equation}
which decays parametrically more slowly than the $1/t$ decay of the isolated point—the reduced
dimensionality of the one-dimensional degenerate stationary manifold is directly reflected in the
difference in the decay exponent ($t^{-1/2}$ vs.\ $t^{-1}$).

To summarize, in the long-time limit the reduced density matrix has the structure
\begin{equation}
\rho_S(t) \;\xrightarrow{\ t\to\infty\ }\;
\rho_S^{(\mathrm{pointer})}
\;+\;
\Big[\, \mathcal{A}\, t^{-1/2} e^{-i(2gt+\pi/4)}\;
\Pi_-^{(+)}(0)\,\rho_S(0)\,\Pi_-^{(-)}(0) \;+\; \mathrm{h.c.}\Big]
\;+\; O(t^{-1}),
\label{eq:asymptotic}
\end{equation}
where the first term $\rho_S^{(\mathrm{pointer})}$ is the diagonal ensemble of Eq.~(19), and the
second term is the contribution from the resonance manifold derived in this section. From the
general theory of the stationary-phase method, the broader the width of the Gaussian
distribution, the more rapidly the contributions away from the stationary region decay. The
approximate condition is
\begin{equation}
\mathcal{A}\ \text{is non-negligible}\quad\Longleftrightarrow\quad
|\Delta_1-\Delta_2|\ \lesssim\ \max\!\left(\frac{\sigma_1^2}{\Delta_1},\frac{\sigma_2^2}{\Delta_2}\right).
\end{equation}

The physically important point is as follows. For each fixed $(B_1,B_2)$, the central two-spin
system generally precesses (undergoes coherent oscillation) at a frequency $2E_-(B_1,B_2)$ that
varies from one environmental configuration to another. Ordinarily, this spread in frequency
leads, upon averaging over the environment, to rapid (Gaussian or exponential) dephasing. On the
resonance manifold $\Omega_1(B_1)=\Omega_2(B_2)$, however, as seen in Eqs.~(27)–(28), the
eigenvalue splitting is constant, independent of $\Omega_r$. Consequently, whichever
environmental configuration on this manifold is chosen, the central system oscillates between the
pair of two-body entangled states $|\chi_-^{(\pm)}(0)\rangle$ at the \emph{same} frequency $2g$,
so that there is no ``relative spread in precession'' among the environmental configurations. In
this sense, the component projected onto the resonance manifold escapes the usual
environment-averaged dephasing and survives as coherence between a pair of protected two-body
entangled states, subject only to the slow (non-exponential) $t^{-1/2}$ decay.

\section{The Quantum States Responsible for the Slow Decay}

The slow $t^{-1/2}$ decay derived in the previous section originates from the coherence spanned
by the pair of projectors $\Pi_-^{(+)},\Pi_-^{(-)}$. In this section, we clarify explicitly which
quantum states these projectors correspond to.

\subsection{Effective Two-Level Hamiltonian of the $\Pi=-1$ Block}

In the rotated coordinate frame, $\sigma_y^{(1)}$ and $\sigma_y^{(2)}$ remain invariant under the
local rotations (rotations about the $y$ axis), so the coupling term
$g\,\sigma_y^{(1)}\sigma_y^{(2)}$ retains its original form. The $\Pi=-1$ subspace is spanned by
the two states
\begin{equation}
  \ket{+,-} \equiv \ket{\xi{+}}\otimes\ket{\zeta{-}}, \qquad
  \ket{-,+} \equiv \ket{\xi{-}}\otimes\ket{\zeta{+}}
\end{equation}
(where $\xi,\zeta$ are the rotated quantization axes defined in Eqs.~(10) and (11)). Evaluating
$\sigma_y^{(1)}\sigma_y^{(2)}$ within this subspace, using the relation
$\sigma_y\ket{\pm}=\pm i\ket{\mp}$, the corresponding block reduces to the effective two-level
(pseudospin $\tau$) Hamiltonian
\begin{equation}
  H_- = u\,\tau_z - g\,\tau_x, \qquad u \equiv \Omega_1-\Omega_2,
\end{equation}
where the eigenstates of $\tau_z$ are $\ket{+,-}$ and $\ket{-,+}$. This precisely reproduces
$E_-=\sqrt{u^2+g^2}$ of Eq.~(14).

Diagonalizing this Hamiltonian gives
\begin{equation}
  \ket{E_-^{(\pm)}} = \cos\frac{\alpha}{2}\,\ket{+,-} \mp \sin\frac{\alpha}{2}\,\ket{-,+},
  \qquad \tan\alpha = \frac{g}{u}.
\end{equation}
The mixing angle $\alpha$ is a parameter that interpolates between the product of local energy
eigenstates ($\alpha\to 0$, $u\gg g$) and the maximally entangled state ($\alpha\to \pi/2$).

\subsection{States on the Stationary Curve $u=0$}

Precisely at the stationarity condition $\Omega_1(B_1)=\Omega_2(B_2)$, i.e., at $u=0$, the mixing
angle is fixed at $\alpha=\pi/2$, and
\begin{equation}
  \ket{E_-^{(\pm)}} \;\xrightarrow{u\to 0}\;
  \frac{1}{\sqrt{2}}\Big(\ket{\xi{+},\zeta{-}} \mp \ket{\xi{-},\zeta{+}}\Big),
\end{equation}
an equal superposition of Bell-state type.

The term
\begin{equation}
  \Pi_-^{(+)}\rho_S(0)\Pi_-^{(-)}
\end{equation}
appearing in Eq.~\eqref{eq:asymptotic} is precisely the projector representing the coherence
(off-diagonal element) between the two Bell-type states $\ket{E_-^{(+)}}$ and $\ket{E_-^{(-)}}$,
corresponding to $ g\sigma_y^{(1)}\sigma_y^{(2)} = \mp g$. The integral $\mathcal{A}$ 
quantifies the statistical weight of the environmental configurations satisfying $\Omega_1=\Omega_2$ are''; the quantum state itself that contributes always converges,
regardless of the environmental configuration, to this same pair of $45^{\circ}$-mixed Bell
states. We may thus summarize this as \emph{coherence carried by a pair of maximally entangled,
Bell-type states of the two-spin system, appearing only under the condition that the local fields
cancel each other}.

The physical picture can be summarized as follows.

\begin{itemize}
  \item The $t^{-1}$ decay term originating from the \textbf{isolated stationary point
        ($B_1=B_2=0$)} is coherence between energy eigenstates in the ordinary sense—an
        ``unremarkable'' coherence that decays as rapidly as usual due to environmental noise.
  \item The $t^{-1/2}$ decay term originating from the \textbf{hyperbolic stationary curve
        ($\Omega_1=\Omega_2$)} is coherence between two states that, although the mixing angle
        differs from one environmental configuration to another, always share the same 
        maximally entangled Bell-type states; it is precisely this
        degeneracy (the entire one-dimensional manifold being stationary) that slows the decay.
\end{itemize}

\section{Discussion}

\subsection{Comparison with Pointer-Basis Theory and Physical Interpretation}

In the original pure-dephasing model~\cite{zurek1981}, since the central spin has no
self-Hamiltonian ($H_S = 0$), the eigenstates of $\sigma_z$—determined by the commutation
relation $[H_I, \sigma_z] = 0$ with the interaction term—were selected as the fixed pointer
basis. In the 2005 analysis, a self-Hamiltonian ($H_S = \Delta \sigma_x$) was introduced, and
the crossover between the self-dynamics and the environmental coupling was
investigated~\cite{zurek2005}. The model of the present study can be regarded as an extension
of this structure—``a spin with a self-Hamiltonian (driving field) coupled to an
environment''—to a two-body system (a dimer).

In ordinary einselection, the pointer basis is determined almost trivially once the
system–environment coupling operator $H_{int}$ is specified (for example, if one writes the
coupling in the form $\sigma_z^{(S)}\otimes\sum_k g_k\sigma_z^{(k)}$, it can be read off
directly from its construction that the pointer basis is given by the eigenstates of
$\sigma_z^{(S)}$).

In contrast, in the present model, the coupling of each spin to its environment is strictly
\emph{local} ($\sigma_z^{(1)}\otimes\sum_k g_k\sigma_z^{(k)}$ and
$\sigma_{z'}^{(2)}\otimes\sum_k g'_k\sigma_{z'}^{(k)}$). Naively, based on this local coupling
alone, one would expect the pointer basis to be the product of the local eigenbases of each
spin. However, what the stationary-phase analysis revealed is that what actually survives over
long times is the eigenbasis of the \emph{inter-system interaction} $g\sigma_y^{(1)}\sigma_y^{(2)}$—that
is, the eigenstates of an operator that does not appear directly in either system–environment
coupling operator (Bell-state-like entangled states).

What governs the selection of this pointer observable is not ``which operator is coupled to
the environment,'' but rather \emph{a statistical matching condition on the environment
side}—namely, ``how resonant the environment is'' ($\Omega_1(B_1)=\Omega_2(B_2)$). This arises
as an effect of ensemble phase-locking: because the energy splitting
$E_-(u)=\sqrt{u^2+g^2}$ has an extremum with respect to the detuning $u=\Omega_1-\Omega_2$ (the
bottom of a quadratic, $dE_-/du|_{u=0}=0$), configurations in the vicinity of the resonance
point remain in phase with one another over long times. This protection, arising from the
``vanishing of first-order sensitivity,'' is mathematically analogous to the mechanism
underlying clock transitions in atomic clocks and sweet-spot operation of superconducting
qubits (in which first-order dephasing vanishes because the energy is stationary with respect
to the noise parameter). It is a weaker mechanism than Zurek's exact commutativity, but one
that appears robustly in more general models (Appendix A).

To summarize,
\begin{center}
\renewcommand{\arraystretch}{1}
\setlength{\tabcolsep}{12pt}
\begin{tabular}{lll}
 & \shortstack{single system\\(Zurek)} & \shortstack{two-body system\\(This work)} \\
\hline
\shortstack{What determines\\the pointer basis} &
\shortstack{System--environment\\coupling operator\\(local, direct)} &
\shortstack{System--system\\coupling operator\\(global)} \\
Selection condition &
\shortstack{Always satisfied\\(trivial)} &
\shortstack{Statistical matching\\between environments\\(resonance)} \\
\end{tabular}
\end{center}

\subsection{Experimental and Practical Significance}

The lower bound of the observable time window is the apparent pure-dephasing time $T_2^*$ due
to static inhomogeneous broadening. $T_2^*$ corresponds to the effective timescale of the
$1/t$ decay arising from the isolated stationary point (the ``ordinary'' configurations away
from resonance); once the bulk of the ensemble loses coherence on this timescale, only the
small subset of configurations near the resonance manifold (with weight $\mathcal{O}$) is
expected to remain as the dominant component of the signal.

On the other hand, the upper bound is set by $T_1$, as the limit of validity of the model:
since the present model treats only adiabatic, purely unitary evolution, it does not in
principle include $T_1$-type relaxation involving actual energy exchange. That is, the
observable window is
\begin{equation}
\frac{1}{g} \ll t \ll T_1,\qquad\text{and}\qquad T_2^* \ll t,
\end{equation}
and the effective width of this window can be estimated by the ratio $T_1/T_2^*$. In
solid-state quantum devices, it is common to have $T_2^*\ll T_1$ (typically by orders of
magnitude), so this window is often wide open:

The conclusions obtained from this model offer the following implications for practical
applications in quantum technology.

\begin{enumerate}
    \item \textbf{Separation of timescales ($T_2^* \ll t \ll T_1$):}
    This analysis accurately describes the dynamical behavior in the intermediate-time regime
    $T_2^* \ll T_1$, predicting that, even after the local superposition has been destroyed,
    the two-body-correlated pointer state survives as a long-lived plateau.
    \item \textbf{Construction of a passive quantum memory:}
    Without relying on active external pulse control (such as dynamical decoupling), simply
    designing the resonance condition of the driving fields $\Delta_1, \Delta_2$ (i.e.,
    $\Delta_1 \approx \Delta_2$) makes the first-order susceptibility to noise vanish,
    allowing a decoupled subspace robust against local noise (the two-body pointer subspace)
    to be geometrically maintained.
\end{enumerate}

\subsection{Outlook and Future Work}
The following analyses remain as topics for future work.

\begin{enumerate}
\item \textbf{Connection to Quantum Darwinism}
The dynamics of microscopic models and information-theoretic criteria ought to be consistent
with one another. In this context, solvable yet realistic models have an important role to
play. Analyzing information-theoretic quantities within the model treated in this paper is
the next task to be addressed.

\item \textbf{Extension to $N\ge 3$}
The state with maximized two-body entanglement was found to be the long-lived quantum state
on the stationary manifold. For example, when the number of mutually interacting central
spins is extended to $N\ge 3$, it is inferred—by the monogamy of quantum entanglement—that
maximal two-body entanglement will again form between some pair among them, while the
relations with the remaining spins reduce to product states. Confirming this naive
expectation is the next task.
\end{enumerate}

\section*{Acknowledgments}
During the preparation of this work, the author used Claude Sonnet 5 and Gemini 3.1 Pro 
in order to assist with English translation, refinement of mathematical exposition, and 
improvement of language readability. Following the use of these tools, the author 
thoroughly reviewed and edited the content as necessary and takes full responsibility 
for the final output of this publication.

\appendix
\section{Generalization: Degenerate Stationary Manifolds and Effective Dimension}

The exact conservation of $\Pi$ arises from a symmetry specific to the present model (an
in-plane $xz$ field combined with a $\sigma_y\sigma_y$ coupling); however, the essential
mechanism responsible for the $1/\sqrt{t}$ decay is more general.

\paragraph{The nature of the mechanism: dependence of the level splitting on the detuning}
Although $E_-(B_1,B_2)=\sqrt{(\Omega_1(B_1)-\Omega_2(B_2))^2+g^2}$ is formally a function of
the two independent variables $B_1,B_2$, it in fact depends on them only through the single
composite variable (the detuning) $u\equiv\Omega_1(B_1)-\Omega_2(B_2)$. By the chain rule,
\begin{equation}
\frac{\partial E_-}{\partial B_1}=\frac{dE_-}{du}\,\frac{d\Omega_1}{dB_1},\qquad
\frac{\partial E_-}{\partial B_2}=-\frac{dE_-}{du}\,\frac{d\Omega_2}{dB_2},
\end{equation}
and since $dE_-/du=u/\sqrt{u^2+g^2}$ vanishes identically at $u=0$ (and \emph{only} at that
point), both partial derivatives vanish simultaneously over the entire level set $u=0$ (i.e.,
the resonance manifold), regardless of the values of $d\Omega_1/dB_1$ and $d\Omega_2/dB_2$.
This is a structure common to avoided crossings (level repulsion) arising from $g\neq0$, and
does not require global exact solvability (the existence of $\Pi$)—it suffices for it to hold
locally within a two-level approximation.

\paragraph{Generalization via dimensional counting}
Stated more abstractly: suppose there are $N$ environmental parameters $B_1,\dots,B_N$, and
the relevant energy splitting depends effectively only on $K<N$ composite variables
($u_1,\dots,u_K$, each a ``detuning''-like quantity), i.e., $E(u_1,\dots,u_K)$. Then the level
set
\begin{equation}
\mathcal{M} = \{(B_1,\dots,B_N) : u_i(B_1,\dots,B_N)=u_i^*,\ i=1,\dots,K\}
\end{equation}
corresponding to a stationary point $u^*=(u_1^*,\dots,u_K^*)$ of $E$ (within the
composite-variable space) is lifted to an $(N-K)$-dimensional stationary manifold along the
remaining $N-K$ ``inert'' directions. By the standard dimensional counting of
multidimensional stationary-phase analysis (only the transverse Fresnel integrals carry
$t$-dependence, while the degenerate directions remain as a $t$-independent weight integral),
the contribution of the coherence term from this manifold decays as
\begin{equation}
\boxed{\ \sim t^{-d_{\rm eff}/2},\qquad d_{\rm eff}\equiv N-\dim\mathcal{M} = K\ }.
\end{equation}
In the present model, $N=2$ ($B_1,B_2$) and the composite variable count is $K=1$
($u=\Omega_1-\Omega_2$), so $d_{\rm eff}=1$, i.e., $t^{-1/2}$. By contrast, the isolated
stationary point (e.g., $B_1=B_2=0$) has both $u$ and $v$ ($\equiv\Omega_1+\Omega_2$) acting
independently, so effectively $K=2=N$, $d_{\rm eff}=2$, giving $t^{-1}$. The two example
calculations of the previous sections correspond to the special cases $K=1,2$ of this general
formula.

\paragraph{Expected generalization}
From this perspective, the following can be expected:
\begin{itemize}
\item For a number of environmental channels $N$, in models where the energy splitting of the
transition of interest depends only on $K$ composite detunings, an $(N-K)$-dimensional
resonance manifold arises, and the corresponding coherence term decays as $t^{-K/2}$.
\item The larger the number of ``inert'' environmental degrees of freedom ($N-K$), the
longer-lived the coherence on the corresponding resonance manifold. If $N=K$ (all channels
contribute to the splitting), only the usual isolated-point-type decay $t^{-N/2}$ occurs, and
no degenerate manifold exists.
\item Conversely, in models where $N$ is increased while $K=1$ is kept fixed (i.e., multiple
environmental degrees of freedom contribute degenerately to the same detuning), $d_{\rm eff}=1$
remains unchanged, and the slow decay $t^{-1/2}$ is robustly preserved—this means that it is
not the ``number'' of environmental degrees of freedom, but rather the ``number of independent
composite variables'' that effectively enter the energy splitting, that determines the decay
exponent.
\end{itemize}
The $\Pi$ symmetry of the present model provides one of the few concrete examples in which
this general structure can be confirmed in closed form as an exact solution.

\end{document}